\documentclass[floatfix,aps,showpacs,twocolumn,nofootinbib,preprintnumbers]
{revtex4}
\usepackage{graphicx}

\usepackage{dcolumn}
\usepackage{bm}

\begin{document}

\def\simlt{\stackrel{<}{{}_\sim}}
\def\simgt{\stackrel{>}{{}_\sim}}

\newcommand{\Frac}[2]{\frac{{\displaystyle #1}}{{\displaystyle #2}}}

\preprint{FERMILAB-Pub-03/212-A}

\title{On the reheating stage after inflation}

\author{Edward W. Kolb}
\affiliation{Fermilab Astrophysics Center, Fermi National
Accelerator Laboratory, Batavia, Illinois, 60510-0500, USA, \\
and Department of Astronomy and Astrophysics, Enrico Fermi
Institute, \\ University of Chicago, Chicago, Illinois 60637-1433,
USA}

\author{Alessio Notari}
\affiliation{Scuola Normale Superiore, Piazza dei Cavalieri 7,  
Pisa I-56126, Italy}

\author{Antonio Riotto}
\affiliation{INFN, Sezione di Padova, Via Marzolo 8, Padova I-35131, Italy}

\date{\today}

\begin{abstract}
\noindent
We point out that inflaton decay products acquire plasma masses during
the reheating phase following inflation.  The plasma masses may render
inflaton decay kinematicaly forbidden, causing the temperature to
remain frozen for a period at a plateau value. We show that the final
reheating temperature may be uniquely determined by the inflaton mass,
and may not depend on its coupling. Our findings have important
implications for the thermal production of dangerous relics during
reheating ({\em e.g.,} gravitinos), for extracting bounds on particle
physics models of inflation from Cosmic Microwave Background
anisotropy data, for the production of massive dark matter candidates
during reheating, and for models of baryogenesis or leptogensis where
massive particles are produced during reheating.

\end{abstract}


\maketitle

%%%%%%%%%%%%%%%%%%%%%%%%%%%%%%%%%%%%%%%%%%%%%%%%%%%%%%%%%%%%%%%%%%
%%%%%%%%%%%%%%%%%%%%%%%%%%%%%%%%%%%%%%%%%%%%%%%%%%%%%%%%%%%%%%%%%%
\section{Introduction}
%%%%%%%%%%%%%%%%%%%%%%%%%%%%%%%%%%%%%%%%%%%%%%%%%%%%%%%%%%%%%%%%%%
%%%%%%%%%%%%%%%%%%%%%%%%%%%%%%%%%%%%%%%%%%%%%%%%%%%%%%%%%%%%%%%%%%

At the end of inflation \cite{review} the energy density of the
universe is locked up in a combination of kinetic energy and potential
energy of the inflaton field $\phi$, with the bulk of the inflaton energy
density in the zero-momentum mode of the field.  Thus, the universe at
the end of inflation is in a cold, low-entropy state with few degrees
of freedom, very much unlike the present hot, high-entropy universe.
After inflation the frozen inflaton-dominated universe must somehow be
defrosted and become a high-entropy, radiation-dominated universe.

One path to defrosting the universe after inflation is known as ``reheating''
\cite{book}. The simplest way to envision the reheating process is if the
comoving energy density in the zero mode of the inflaton decays into normal
particles in a perturbative way. The decay products then scatter and thermalize
to form a thermal background.\footnote{We do not consider here the possible
role of nonlinear dynamics leading to explosive particle production known as
``preheating'' \cite{pre}.}

Of particular interest is a quantity known as the reheat temperature, denoted
as $T_{RH}$. The reheat temperature is properly thought of as the maximum
temperature of the radiation-dominated universe.  It is not necessarily the
maximum temperature obtained by the universe after inflation \cite{book,st,
ckr, gkr}.

The reheat temperature is defined by assuming an instantaneous conversion of
the energy density in the inflaton field into radiation when the decay width of
the inflaton energy, $\Gamma_\phi$, is equal to $H$, the expansion rate of the
universe.  The reheat temperature is calculated quite easily \cite{book}.
After inflation the inflaton field executes coherent oscillations about the
minimum of the potential.  Averaged over several oscillations, the coherent
oscillation energy density redshifts as matter: $\rho_\phi \propto a^{-3}$,
where $a$ is the Robertson--Walker scale factor.  If we denote as $\rho_I$ and
$a_I$ the total inflaton energy density and the scale factor at the onset of
coherent oscillations immediately after the end of inflation, then the Hubble
expansion rate as a function of $a$ is ($M_{Pl}$ is the Planck mass)
\begin{equation}
H(a) = \sqrt{\frac{8\pi}{3}\frac{\rho_I}{M^2_{Pl}}
	\left( \frac{a_I}{a} \right)^3}\ .
\end{equation}
Equating $H(a)$ and $\Gamma_\phi$ leads to an expression for $a_I/a$.
Now if we assume that all available coherent energy density is
instantaneously converted into radiation at this value of $a_I/a$, we
can define the reheat temperature by setting the coherent energy
density, $\rho_\phi=\rho_I(a_I/a)^3$, equal to the radiation energy
density, $\rho_R=(\pi^2/30)g_*T_{RH}^4$, where $g_*$ is the effective
number of relativistic degrees of freedom at temperature $T_{RH}$.
The result is
\begin{eqnarray}
T_{RH} & = & \left( \frac{90}{8\pi^3g_*} \right)^{1/4}
		\alpha_{\phi} ^{1/2} \sqrt{ M_\phi M_{Pl} } \nonumber \\
       & = & 0.2 \left(\frac{100}{g_*}\right)^{1/4}
	     \alpha_{\phi} ^{1/2}  \sqrt{ M_\phi M_{Pl} } \ ,
\label{eq:trh2}
\end{eqnarray}
where we have expressed the inflaton decay width as $\Gamma_\phi=\alpha_\phi
M_\phi$. 

There are various reasons to suspect that the reheating temperature is
small. For instance, in local supersymmetric theories \cite{sugra} gravitinos
(and other dangerous relics like moduli fields) are produced during reheating.
Unless reheating is delayed, gravitinos will be overproduced, leading to a
large undesired entropy production when they decay after big-bang
nucleosynthesis \cite{ellis}. The limit from gravitino overproduction is
$T_{RH} \simlt 10^{9}$ to $10^{10}$GeV, or even stronger \cite{nucleo}.

Again, we emphasize that the reheat temperature is best regarded as the
temperature below which the universe expands as a radiation-dominated universe,
with the scale factor decreasing as $g_*^{-1/3}T^{-1}$.  In this regard it has
a limited meaning \cite{book,st}.  As the scalar field decays into light
states, the decay products rapidly thermalize forming a plasma with temperature
$T$. The latter grows until it reaches a maximum value $T_{MAX}$ and then
decreases as $T\propto a^{-3/8}$ down to the temperature $T_{RH}$, which {\em
should not} be used as the maximum temperature obtained by the universe during
reheating.  The maximum temperature is, in fact, much larger than $T_{RH}$ and
it is incorrect to assume that the maximum abundance of a massive particle
species produced after inflation is suppressed by a factor of
$\exp(-M/T_{RH})$. This has important implications for the idea of superheavy
dark matter \cite{ckr}, supersymmetric dark matter \cite{gkr,sdm} and
baryogenesis \cite{baryo}.

The goal of this paper is to present a simple, but relevant observation that
changes the usual picture of the temperature evolution during reheating.
During the process of reheating the inflaton decay products scatter and
thermalize to form a thermal background. A thermalized particle species
produced during the first stages of reheating acquires a plasma mass $m_p(T)$
of the order of $gT$, where $g$ is the typical (gauge) coupling governing the
particle interactions \cite{weldon}. This happens because forward scatterings
of fermions do not change the distribution functions of particles, but modify
their free dispersion relations, producing a plasma mass. The dispersion
relation can be well-approximated for both scalars and fermions by
$\omega^2=k^2 + m_p^2(T)$, where $\omega$ and $k$ are the energy and the
three-momentum of the particle in the thermal background, respectively.  The
presence of thermal masses imply that the inflaton zero-mode cannot decay into
light states if its mass $M_\phi$ is smaller than about $gT$. The decay process
is simply kinematically forbidden.\footnote{This observation was made first in
in the context of the Affleck-Dine baryogenesis scenario \cite{andrei}.}

Our observation is that during the reheating stage, the inflaton starts
decaying and the temperature of the plasma rises. If the maximum temperature
obtained by the universe during reheating, $T_{MAX}$, is larger than about
$g^{-1}M_\phi$, the inflaton decay channel into light states become
inaccessible and the decay process stops as soon as the temperature has reached
a value of the order of $g^{-1}M_\phi$. Subsequently, expansion cools the
plasma, lowering the temperature and the corresponding plasma masses of the
light states. The inflaton is the free to decay. However, as soon as this
happens, the temperature of the plasma rises and the inflaton decay process
becomes kinematically forbidden again. As a result, one expects a prolonged
period during which the temperature of the plasma is frozen to a plateau value
of the order of $g^{-1}M_\phi$.

Our observation has various implications. First of all, let us notice that we
do not know the mass of the inflaton field around the minimum of its potential
during the reheating stage. Indeed, from the recent WMAP cosmic microwave
background (CMB) anisotropy data \cite{wmap} we only have limited informations
about that portion of the inflaton potential experienced by the inflaton field
during inflation; we know that it has to be quite flat in order to allow a
sufficiently long period of exponential growth of the scale factor
\cite{review,barger,us,leach}. However, we know nothing about the inflaton 
mass during reheating, since this depends upon a portion of the inflaton
potential which is not accessible to any observations. This amounts to saying
that $M_\phi$ should be regarded as a free parameter. Even more, in many
inflationary scenarios, {\it e.g.}, hybrid models \cite{review}, the reheating
dynamics may be determined by a scalar field $\chi$ different from the inflaton
field. (In the following, the terminology ``mass of the inflaton'' will be
therefore used in a loose way.)

Suppose that the reheating temperature $T_{RH}$ defined in Eq.\
(\ref{eq:trh2}), is larger than $g^{-1}M_\phi$. This means that when
the inflaton decay lifetime is of the order of the age of the
universe, the inflaton field would like to decay, but is not allowed
to because the plasma masses of the light decay products are too
large. Only when the energy density stored in the inflaton field
becomes smaller than about $\rho_\phi\sim (g^{-1}M_\phi)^4$ will the
particles in the plasma have a mass smaller than $M_\phi$ and inflaton
can promptly decay.  Under these circumstances the reheating
temperature of the universe should be
\begin{equation}
T_{RH} \simeq \frac{M_\phi}{g},
\label{new}
\end{equation}
which is directly related to the inflaton mass and independent of the
inflaton decay rate! 

Before concluding the introduction, we note that our effect is applicable in
situations other than reheating after inflation.  It would apply, for instance,
if the universe is ever dominated by a decaying nonrelativistic particle.

The rest of the paper is organized as follows. In Sec.\ II we analyze in detail
the behavior of the temperature during the reheating stage, and in particular
we characterize the plateau stage both analytically and numerically. Section
III is devoted to the study of some applications of our findings. We focus on
the production of gravitinos during reheating, on the evaluation of the number
of e-folds after inflation which has recently acquired particular relevance in
order to restrict models of inflation from the CMB anisotropy data, and on the
production of massive particles.  Finally, in Sec.\ IV we present our
conclusions.

%%%%%%%%%%%%%%%%%%%%%%%%%%%%%%%%%%%%%%%%%%%%%%%%%%%%%%%%%%%%%%%%%%
\section{REHEATING WITH THERMAL MASSES}
%%%%%%%%%%%%%%%%%%%%%%%%%%%%%%%%%%%%%%%%%%%%%%%%%%%%%%%%%%%%%%%%%%

We now discuss the reheating process, assuming that the decay products of the
inflaton field rapidly thermalize and acquire ``plasma'' masses $m_p(T)$ of
order $gT$, where $g$ is the coupling constant for a particle in the
plasma.\footnote{We expect $g \lesssim 1$, and will assume $g=1/2$ for
numerical estimates.}

There are two assumptions that deserve elaboration. The first aspect is the
assumption of ``rapid'' thermalization.  The timescale for thermalization of
the inflaton decay products is $(n\sigma)^{-1}$ where $\sigma$ is a cross
section for the scattering of the decay products and $n$ is the number density
of scatterers.  The thermalization is rapid if this timescale is short compared
to the timescale for energy extraction from the inflaton, assumed to be equal
to the lifetime of the inflaton, $\Gamma_\phi^{-1}$. It is reasonable to assume
that the inflaton is weakly coupled and the inflaton lifetime is large compared
to the thermalization time; hence rapid thermalization \cite{ckr}.

The second important aspect of the assumption is that the inflaton decay
products have a thermal mass of order $gT$, where $g\sim 0.5$ is a typical
gauge coupling constant.  One might imagine that the inflaton decays into some
weakly-interacting particles which then subsequently decay into ``thermal''
particles with gauge interactions.  But in any case, eventually the decay
sequence must include particles with gauge interactions for which there will be
a thermal mass.

To model the effect of plasma masses, let us consider, for the moment, a model
universe with two components: inflaton field energy, $\rho_\phi$, and radiation
energy density, $\rho_R$, which contains all the light degrees of freedom
produced after decay.  For simplicity, we can think that the produced particles
in the radiation component all have couplings of the same
strength.\footnote{Since the energies are well above the weak scale, even
neutrinos will interact with gauge-coupling strength.} Also, we consider the
simplest type of decay, that is the decay of the inflaton into scalars.  In the
case of decay into scalars, the only effect of the masses is to modify the
phase space of the products, while the case of fermions is slightly different,
since the scattering amplitude also depends on the masses.

The presence of thermal masses imply that the decay width of the inflaton is no
longer the zero-temperature result $\Gamma_{\phi}=\alpha_{\phi} M_{\phi}$, but
becomes
\begin{equation}
\label{gamma}
\Gamma_\phi(T) = \alpha_\phi M_\phi \sqrt{1-4 \frac{m^2_p(T)}{M^2_{\phi}}}=
             \alpha_\phi M_\phi \sqrt{1-4 \frac{g^2 T^2}{M^2_{\phi}}}   \ .
\end{equation}
The consequence of this simple fact is that the dynamics of reheating
drastically changes when the temperature of the plasma is such that $m_p(T)$
becomes as large as $M_{\phi}$.  When $m_p(T)\ll M_{\phi}$, the effect is
negligible, while the decays stop when $m_p(T)\approx M_{\phi}$ since the
phase space factor goes to zero as $T=M_{\phi}/2 g\approx M_{\phi}$.

With the above assumptions, the Boltzmann equations describing the
redshift and interchange in the energy density among the different
components are
\begin{eqnarray}
\label{eq:BOLTZMANN}
& &\dot{\rho}_\phi + 3H\rho_\phi +\Gamma_\phi(T) \rho_\phi = 0
	\nonumber \\
& & \dot{\rho}_R + 4H\rho_R - \Gamma_\phi(T) \rho_\phi = 0
  \ ,
\end{eqnarray}
where dot denotes time derivative.

It is clear that the system behaves in such a way that $T$ never becomes larger
than $M_{\phi}/2 g$, otherwise the factor $\Gamma_{\phi}(T)$ would become
imaginary.  In other words, when $T$ reaches this value we have a phase with
approximately constant $T$ during which the decays are suppressed for kinematic
reasons.  During this phase $\rho_R$ stays constant, while $\rho_{\phi}$
decreases like $a^{-3}$.  We recall that without plasma masses, the behavior of
$T$ is very different: immediately after inflation ends it grows rapidly to
$T_{MAX}$, and then decreases like $a^{-3/8}$ until it reaches $T_{RH}$.  At
this point the $\phi$ field decays completely and the universe becomes
radiation dominated.

Taking into account the effect of plasma masses, we may have three
possibilities:
\[
\begin{array}{rrcl}
\textrm{I:}   \hspace*{0.5cm} & T_{MAX} < & M_{\phi} &           \\
\textrm{II:}  \hspace*{0.5cm} & T_{RH} <  & M_{\phi} & < T_{MAX} \\
\textrm{III:} \hspace*{0.5cm} &           & M_{\phi} & <  T_{RH}.
\end{array}
\]
In case I, the effect of the plasma mass is negligible. In case II,
after a very short time $T$ grows to $M_{\phi}$, then stays approximately
constant for a while, then decreases as $a^{-3/8}$ until reheating and the
radiation dominated phase begins. In case III, again after a very short
time, $T$ grows to $M_{\phi}$ and after a long phase of constant $T$, the
universe directly enters the radiation-dominated phase after the time of
reheating determined ignoring plasma effects.

We want now to discriminate, in terms of the fundamental parameters of the
inflaton field, the applicable case (I, II, or III), and the duration of the
constant-$T$ phase.  First, recall that the maximum temperature obtained after
inflation is given by
\cite{gkr}
\begin{eqnarray} 
T_{MAX} & = & \left( \frac{3}{8} \right)^{2/5} 
		\left( \frac{15}{2\pi^3} \right)^{1/4}
\alpha_\phi^{1/4}\left(\frac{M_{Pl}^2H_I}{g_*M_\phi^3}\right)^{1/4} M_\phi 
       \nonumber \\ 
        & = & 0.6
\alpha_\phi^{1/4}\left(\frac{M_{Pl}V^{1/2}}{g_*M_\phi^3}\right)^{1/4} M_\phi\ ,
\end{eqnarray}
where $V$ is the value of the inflaton potential at the end of inflation.  The
reheating temperature was defined in Eq.\ (\ref{eq:trh2}).  We may now
determine the conditions that determine the operative case in terms of the
value of the decay constant $\alpha_{\phi}$:
\[
\begin{array}{rrcl}
\textrm{I:}   \hspace*{.5cm} & & \alpha_{\phi} & \lesssim  
	g_* \Frac{M_{\phi}^3}{g^4 M_{Pl} V^{1/2}} \\
\textrm{II:}  \hspace*{.5cm}& g_* \Frac{M_{\phi}^3}{g^4 M_{Pl} V^{1/2}} 
								\lesssim &
        \alpha_{\phi} & \lesssim g_*^{1/2} \Frac{M_{\phi}}{g^{2} M_{Pl}}  \\
\textrm{III:} \hspace*{.5cm} & g_*^{1/2} \Frac{M_{\phi}}{g^{2}M_{Pl}}\lesssim 
        & \alpha_{\phi} &.
	\end{array}
\]
We see that, for realistic values of parameters, it is likely that we are in
the second or in the third case, {\it i.e.,} the effect is non-negligible. If
we put $V^{1/4} \approx 10^{13}$GeV, $M_{\phi}\approx 10^8$GeV and
$g_*\approx 10^2$, we obtain
\[
\begin{array}{rrcl}
\textrm{I:}   \hspace*{0.5cm} &  & \alpha_{\phi}& \lesssim 10^{-18} \\
\textrm{II:}  \hspace*{0.5cm} & 10^{-18} \lesssim &\alpha_{\phi}& 
	\lesssim 3\times10^{-10} \\
\textrm{III:} \hspace*{0.5cm} & 3\times10^{-10} \lesssim &\alpha_{\phi}&.
	\end{array}
\]

Next, we may estimate the duration of the constant-$T$ phase in cases II
and III.  We will denote by $a_I$ the value of the scale factor at the
beginning of the reheating phase, and by $a_F$ its value at the end of the
constant-$T$ phase.

In case II, $a_F$ may be estimated by assuming the usual scaling of the
temperature ignoring plasma mass effects during reheating, $T\propto a^{-3/8}$,
and finding the value of $a$ when $T$ drops below the value $M_{\phi}/2 g$.
The behavior of $T$ is
\begin{equation}
\label{threeeights}
\frac{T}{M_\phi}  \simeq  \left(\frac{54}{\pi^5}\right)^{1/8}
	\alpha_{\phi}^{1/4}
        \left( \frac{ M_{Pl} V^{1/2}}{g_*M_{\phi}^3} \right)^{1/4}
        \left(\frac{a}{a_I}\right)^{-3/8} .
\end{equation}
Imposing the condition $T/M_\phi=1/2g$ to define $a_F$ we find
\begin{equation}
\frac{a_F}{a_I} = (2g)^{8/3} \left( \frac{54}{\pi^5}\right)^{1/3}
	\alpha_\phi^{2/3} \frac{ M_{Pl}^{2/3} V^{1/3} } {M_\phi^2} .
      \label{IIaFaI}
\end{equation}
In terms of number of e-folds, imposing $V^{1/4} \approx 
10^{14}$GeV, $M_{\phi}\approx 10^9$GeV) we obtain $N \approx 30 + 2/3
\ln(\alpha_{\phi})$.

In case III, the situation is much different than the case ignoring
plasma effects.  In the usual case (without plasma masses) the system would
enter the radiation-dominated era at the time of $\phi$ decay
($\Gamma_{\phi}=H$):
\begin{equation}
\frac{a_{RH}}{a_I} = \left( \frac{8\pi}{3}\right)^{1/3}
  \frac{V^{1/3}}{M_{\phi}^{2/3} M_{Pl}^{2/3} \alpha_{\phi}^{2/3}} .
\end{equation}
In our case, though, decays are not possible so long as $T$ is larger than
$M_{\phi}/2 g$.  So, the $\phi$ energy density continues evolving approximately
like $a^{-3}$ until $\rho_{\phi}$ becomes smaller than $\rho_R$, at which time
the $\phi$ can decay without enhancing the temperature (and so closing the
phase space for the decay).  So the condition is simply for case III is
\begin{equation}
V\left(\frac{a_I}{a_F}\right)^3 \lesssim \frac{\pi^2}{30} g_*
 \left( \frac{M_{\phi}}{2 g}\right )^4,
\end{equation}
which implies
\begin{equation}
\frac{a_{F}}{a_I} \approx  4 g^{4/3}
        \left( \frac{V}{g_* M_{\phi}^4} \right)^{1/3}
        \approx \left(\frac{V}{M_\phi^4}\right)^{1/3}.
\label{IIIaFaI}
\end{equation}
In terms of number of e-folds, imposing again realistic values for this case,
$V^{1/4} \approx 8\times10^{11}\textrm{GeV}$, $M_{\phi}\approx
2\times10^7\textrm{GeV}$, we obtain $N \approx 14$.  The two cases reduce to
the same value in the intermediate case ({\it i.e.,} the case in which
$T_{RH}\simeq M_\phi$).

%%%%%%%%%%%%%%%%%%%%%%%%%%%%%%%%%%%%%%%%%%%%%%%%%%%%%%%%%%%%%%%%%%
%%%%%%%%%%%%%%%%%%%%%%%%%%%%%%%%%%%%%%%%%%%%%%%%%%%%%%%%%%%%%%%%%%
\begin{figure}
\includegraphics[width=0.45\textwidth]{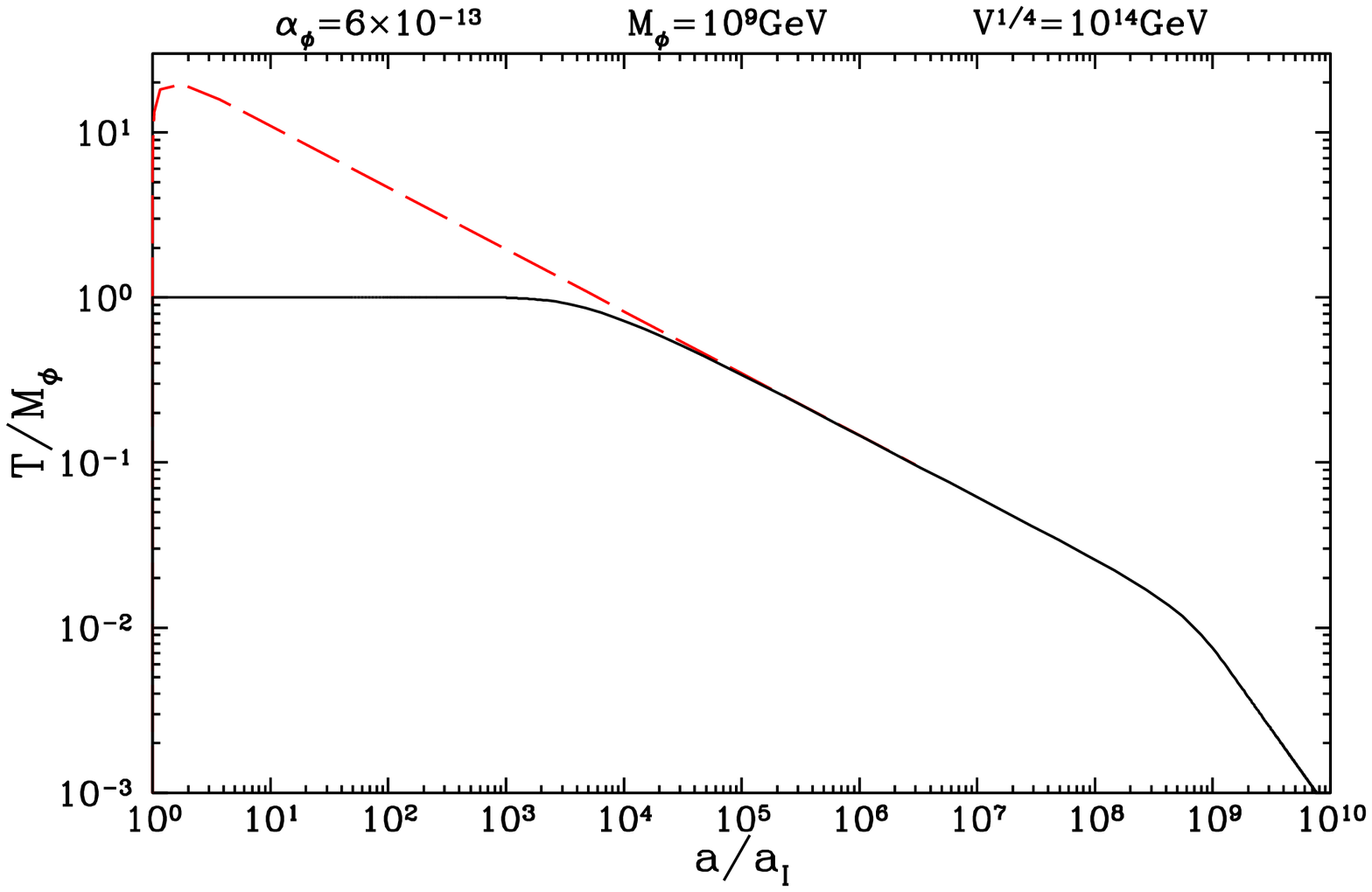}
\caption{\label{fig:caseII}\em 
The behavior of the temperature during 
reheating, without (dashed line) and with (solid line) plasma mass effects,
for case II: $T_{RH} < M_{\phi}$.}
\end{figure}
%%%%%%%%%%%%%%%%%%%%%%%%%%%%%%%%%%%%%%%%%%%%%%%%%%%%%%%%%%%%%%%%%%
%%%%%%%%%%%%%%%%%%%%%%%%%%%%%%%%%%%%%%%%%%%%%%%%%%%%%%%%%%%%%%%%%%

%%%%%%%%%%%%%%%%%%%%%%%%%%%%%%%%%%%%%%%%%%%%%%%%%%%%%%%%%%%%%%%%%%
%%%%%%%%%%%%%%%%%%%%%%%%%%%%%%%%%%%%%%%%%%%%%%%%%%%%%%%%%%%%%%%%%%
\begin{figure}
\includegraphics[width=0.45\textwidth]{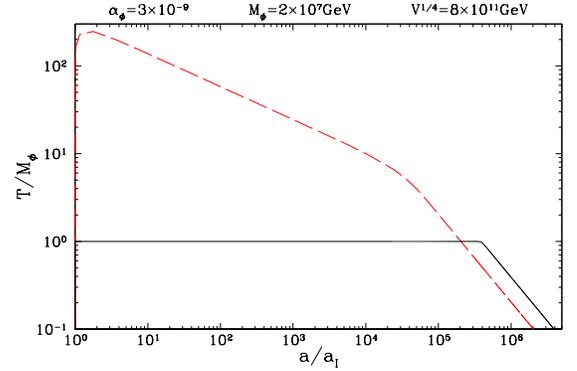}
\caption{\label{fig:caseIII}\em The behavior of the temperature during 
reheating, without (dashed line) and with (solid line) plasma mass effects, 
for case III: $T_{RH} > M_{\phi}$.}
\end{figure}
%%%%%%%%%%%%%%%%%%%%%%%%%%%%%%%%%%%%%%%%%%%%%%%%%%%%%%%%%%%%%%%%%%
%%%%%%%%%%%%%%%%%%%%%%%%%%%%%%%%%%%%%%%%%%%%%%%%%%%%%%%%%%%%%%%%%%

Now we want to analyze in detail what happens to the system in cases II
and III by numerically solving the Boltzmann equations.  In order to do
this it is more convenient to express the Boltzmann equations in terms of
dimensionless quantities that can absorb the effect of expansion of the
universe.  This may be accomplished with the definitions
\begin{equation}
\label{def}
\Phi \equiv \rho_\phi M_\phi^{-1} a^3 \ ; \quad
R    \equiv \rho_R a^4 .
\end{equation}
It is also convenient to use the scale factor, rather than time, as the
independent variable, so we define a variable $x = a M_\phi$.  With this choice
the system of equations can be written as (prime denotes $d/dx$)
\begin{eqnarray}
\label{SYS}
\Phi' & = & -   \sqrt{\frac{3}{8\pi}} \frac{M_{Pl}}{M_\phi}\alpha_\phi
             \sqrt{1-4 \frac{g^2 T^2(x)}{M^2_{\phi}}} 
\frac{x}{\sqrt{\Phi x + R}}   \ \Phi \nonumber \\
R'    & = &    \sqrt{\frac{3}{8\pi}} \frac{M_{Pl}}{M_\phi}\alpha_\phi
\sqrt{1-4 \frac{g^2 T^2(x)}{M^2_{\phi}} } 
\frac{x^2}{\sqrt{\Phi x + R}} \ \Phi ,
\end{eqnarray}
where the temperature $T(x)$ depends upon $R$ and $g_*$, the effective number
of degrees of freedom in the radiation:
\begin{equation}
\label{Temp}
\frac{T(x)}{M_{\phi}} = \left( \frac{30}{g_*\pi^2}\right)^{1/4}
\frac{R^{1/4}}{x} \ .
\end{equation}

It is straightforward to solve the system of equations in Eq.\
(\ref{SYS}) with initial conditions at $x=x_I$ of $R(x_I)=X(x_I)=0$
and $\Phi(x_I)=\Phi_I$.  It is convenient to express
$\rho_\phi(x=x_I)$ in terms of the expansion rate at $x_I$, which
leads to
\begin{equation}
\Phi_I = \frac{3}{8 \pi} \frac{M^2_{Pl}}{M_\phi^2}
		\frac{H_I^2}{M_\phi^2}\ x_I^3 \ .
\end{equation}
The numerical value of $x_I$ is irrelevant.

We show in Figs.\ \ref{fig:caseII} and \ref{fig:caseIII} the solution of the
system respectively in cases II and III. They follow the qualitative behavior
we described, with the prominent constant-$T$ phase.

%%%%%%%%%%%%%%%%%%%%%%%%%%%%%%%%%%%%%%%%%%%%%%%%%%%%%%%%%%%%%%%%%%
%%%%%%%%%%%%%%%%%%%%%%%%%%%%%%%%%%%%%%%%%%%%%%%%%%%%%%%%%%%%%%%%%%
\section{applications}
%%%%%%%%%%%%%%%%%%%%%%%%%%%%%%%%%%%%%%%%%%%%%%%%%%%%%%%%%%%%%%%%%%
%%%%%%%%%%%%%%%%%%%%%%%%%%%%%%%%%%%%%%%%%%%%%%%%%%%%%%%%%%%%%%%%%%

%%%%%%%%%%%%%%%%%%%%%%%%%%%%%%%%%%%%%%%%%%%%%%%%%%%%%%%%%%%%%%%%%%
\subsection{Thermal production of gravitinos}
%%%%%%%%%%%%%%%%%%%%%%%%%%%%%%%%%%%%%%%%%%%%%%%%%%%%%%%%%%%%%%%%%%

The first question we want to address is the production of gravitinos during
reheating, taking into account of the effect of thermal masses.\footnote{Here
we consider gravitinos, but the results are easily generalized to other
dangerous light relics.}  It is known that the overproduction of gravitinos
represents a major obstacle in constructing cosmological models based on
supergravity \cite{sugra}.  Gravitinos decay very late and, if they are
copiously produced during the evolution of the early universe, their energetic
decay products destroy $^4$He and D by photodissociation, thus jeopardizing the
successful nucleosynthesis predictions \cite{nucleo,ellis}. As a consequence,
the ratio of the number density of gravitinos, $n_{3/2}$, to the entropy
density, $s$, should be smaller than about 
\begin{equation}
\label{lll}
\frac{n_{3/2}}{s}\lesssim 10^{-12} ,
\end{equation}
for gravitinos with mass of the order of 100 GeV.

Gravitinos can be produced in the early universe because of thermal scatterings
in the plasma during the stage of reheating after inflation.  Usually, to avoid
the overproduction of gravitinos, one has to require that the reheating
temperature $T_{RH}$ after inflation is not larger than about $10^{8}$ to
$10^{9}$ GeV \cite{ellis}.  In our case, the relevant parameter is no longer
$T_{RH}$, since the temperature is cutoff by the effect of thermal masses.  We
present here an analysis of the thermal generation of gravitinos during
reheating with a phase of constant temperature.

Recall the salient aspects of the calculation of the gravitino abundance
without thermal masses.  The gravitino abundance is determined by the 
Boltzmann equation
\begin{equation}
\label{l}
\frac{d n_{3/2}}{dt}+3H n_{3/2}=-\langle\sigma_Av\rangle 
\left(n_{3/2}^2-(n^2_{3/2})_{eq}\right),
\end{equation}
where $\langle\sigma_A v\rangle \propto 1/M_{Pl}^2$ is the thermal average of
the gravitino annihilation cross section times the M{\o}ller velocity.
Assuming the actual gravitino density is much less than its equilibrium value
$(n_{3/2})_{eq}=3 g_{3/2}\zeta(3)T^3/4\pi^2$ ($g_{3/2}$ is the number of
degrees of freedom of the gravitino), the evolution of the comoving gravitino 
number density ($N=a^3 n_{3/2}$) is quite simple:
\begin{equation}
\label{ltwo}
\frac{d N_{3/2}}{da} = \frac{ca^2T^6}{HM^2_{Pl}} ,
\end{equation}
where $c=(3g_{3/2}\zeta(3)/4\pi^2)^2$. 

In the radiation-dominated phase $H\propto a^{-2}$ and $T\propto a^{-1}$, so
the dominant contribution to $N_{3/2}$ comes from small $a$, corresponding to
large $T$. During reheating $H\propto a^{-3/2}$.  If plasma effects are not
important $T\propto a^{-3/8}$ during reheating, while if plasma effects are
important $T\propto \textrm{const.}$ during reheating.  In either case, the
dominant  contribution to $N_{3/2}$ comes from large $a$, corresponding to
the end of reheating. Therefore we can calculate $N_{3/2}$ at the end of
the reheating era (the beginning of the radiation-dominated era), and compare
it to the comoving entropy density $N_s=a^3 T^3 2\pi^2g_*/45$.  The result is
\begin{equation}
\frac{N_{3/2}}{N_s}=\frac{n_{3/2}}{s} \approx \left\{ 
\begin{array}{lll}
10^{-2}\Frac{T_{RH}}{M_{Pl}} & (T_{RH}<M_\phi & \textrm{cases I, II})\\
10^{-2}\Frac{M_\phi}{M_{Pl}} & (T_{RH}>M_\phi & \textrm{case III}). \\
\end{array} \right.
\label{ll} 
\end{equation}

Comparing Eqs.\ (\ref{lll}) and (\ref{ll}), one obtains the bounds
\begin{equation}
(10^{8}-10^{9}) \, \textrm{GeV} \gtrsim \left\{
\begin{array}{lll}
T_{RH} & (T_{RH}<M_\phi & \textrm{cases I, II})\\
M_\phi  & (T_{RH}>M_\phi & \textrm{case III}).\\
\end{array} \right.
\end{equation}

This calculation illustrates the point that in case III, the reheat
temperature $T_{RH}$ has no meaning.

%%%%%%%%%%%%%%%%%%%%%%%%%%%%%%%%%%%%%%%%%%%%%%%%%%%%%%%%%%%%%%%%%%
\subsection{Number of e-folds after inflation}
%%%%%%%%%%%%%%%%%%%%%%%%%%%%%%%%%%%%%%%%%%%%%%%%%%%%%%%%%%%%%%%%%%

The quality and quantity of observational data has reached the point
where it is possible to start to place meaningful constraints on
inflationary models \cite{wmap,barger,us}.  In the phenomenology of
extracting predictions from even simple inflation models, one of the
significant uncertainties is the location of the inflaton
corresponding to when scales of observational interest crossed the
Hubble radius during inflation.  Recent studies of this issue
\cite{leach,scottlam} have pointed out that a significant factor is
the uncertainty in the duration of the reheating phase.  Lack of
knowledge of the duration of the reheating results in an uncertainty
in the number of e-folds of expansion after inflation ends
\cite{book}.  The uncertainty is usually parameterized in terms of the 
reheat temperature, with the uncertainty in the number of e-folds of
inflation depending on $\ln T_{RH}^{1/3}$.

As we have stressed, in case III the reheat temperature has no
meaning; the radiation-dominated era commences with $T=M_\phi$.  
If case III obtains, then previous formulas for the number of e-folds
should depend on 
\begin{equation}
\Delta N = \frac{1}{3}\ln \frac{M_\phi}{V^{1/4}} ,
\end{equation}
instead of the traditional formula used for $\Delta N$ \cite{book}, 
$\Delta N = \frac{1}{3}\ln T_{RH}/V^{1/4}$, {{\it i.e.,} $T_{RH}\sim 
\sqrt{\Gamma_\phi M_{Pl}}$ should be replaced by $M_\phi$. This means 
that if case {\bf III} is attained, the number of e-folds corresponding to
scales of observational interest is smaller than in the usually adopted case by
a factor $\frac{1}{3}\ln \sqrt{\alpha_\phi M_{Pl}/M_\phi}$.

Proper calculation of the number of e-folds after inflation is crucial
in determining the viability of inflation models.  The change in the
number of e-folds in case III may be crucial.

%%%%%%%%%%%%%%%%%%%%%%%%%%%%%%%%%%%%%%%%%%%%%%%%%%%%%%%%%%%%%%%%%%
\subsection{Production of massive particles}
%%%%%%%%%%%%%%%%%%%%%%%%%%%%%%%%%%%%%%%%%%%%%%%%%%%%%%%%%%%%%%%%%%

Our findings may be relevant for the production of massive particles
during the reheating stage and, in particular, for the production of
superheavy dark matter (WIMPZILLAS) \cite{ckr,shdm} and leptogenesis
\cite{lepto}.

There are many reasons to believe the present mass density of the
universe is dominated by a weakly interacting massive particle (WIMP),
a fossil relic of the early universe. Theoretical ideas and
experimental efforts have focused mostly on production and detection
of thermal relics, with mass typically in the range a few GeV to a
hundred GeV. However, during the transition from the end of inflation
to the beginning of the radiation phase, superheavy and nonthermal
particles may be generated. If they are stable they may provide a
significant contribution to the total dark matter density of the
universe.

Let us consider a superheavy particle $X$ with mass $M_X$. In this
section we will restrict our attention to case {\bf III} for which the
final reheating temperature is fixed by the inflaton mass, and we
consider the case in which $M_X>M_\phi$.\footnote{In the opposite
case, the superheavy dark matter may be generated directly by the
inflaton decay and its abundance will depend on the coupling between
the inflaton and the $X$-particles. In such a case, the final
abundance of the superheavy dark matter will be model dependent.} We
suppose that the $X$-particles are produced in pairs during the
reheating stage by annihilation of light states. The corresponding
Boltzmann equation for the number density $n_X$ reads
\begin{equation}
\frac{dn_X}{dt}+3H n_X=-\langle \sigma_A v\rangle\left(n_X^2 -
\left(n_X^2\right)_{eq}\right),
\end{equation}
where $\langle \sigma_A v\rangle\simeq \alpha_X/M_X^2$ is the thermal
average of the annihilation cross section times the M{\o}ller
velocity.  Assuming the actual density $n_X$ is much less than its
equilibrium value $(n_X)_{eq}= g_{X}(M_X T/2\pi)^{3/2}e^{-M_X/T}$
($g_{X}$ is the number of degrees of freedom of the $X$-particles) and
remembering that dominant contribution to the production comes from
end of reheating when the temperature is of the order of $M_\phi$, we
can estimate the ratio between the number density of $X$-particles and
the entropy density at the end of reheating to be
\begin{equation}
\frac{n_X}{s}\simeq 10^{-2}\frac{g_X^2}{g_*^{3/2}} M_{Pl}M_X
\langle \sigma_A v\rangle \left(\frac{M_X}{M_\phi}\right)^2 e^{-2M_X/M_\phi},
\end{equation}
corresponding to a present-day abundance of
\begin{equation}
\label{omega}
\Omega_X h^2\simeq 10^{22} g_X^2 M_X^2\langle \sigma_A v\rangle\left(
\frac{M_X}{M_\phi}\right)^2 e^{-2M_X/M_\phi}.
\end{equation}
Taking $ M_X^2\langle \sigma_A v\rangle\sim 1$, a moderate hierarchy
between the inflaton mass and the superheavy dark matter particle
$M_X$, $M_X/M_\phi\sim 30$, may explain the observed value for the
dark matter abundance of about 30\%.  Eq.\ (\ref{omega}) is much
different than previous results \cite{ckr}.

Our findings also have important implications for the conjecture that
ultra-high energy cosmic rays, above the Greisen-Zatsepin-Kuzmin
cutoff of the cosmic ray spectrum, may be produced in decays of
superheavy long-living particles \cite{kr1,kr2,bs}. In order to produce
cosmic rays of energies larger than about $10^{13}$ GeV, the mass of
the $X$-particles must be very large, $M_X\simgt 10^{13}$GeV, and
their lifetime $\tau_X$ cannot be much smaller than the age of the
Universe, $\tau_X\simgt 10^{10}$ yr.  With the smallest value of the
lifetime, the observed flux of ultra-high energy cosmic rays will be
reproduced with a rather low density of $X$-particles, $\Omega_X\sim
10^{-12}$. The expression Eq.\ (\ref{omega}) suggests that the
$X$-particles can be produced in the right amount by collisions taking
place during the reheating stage after inflation if the inflaton mass
is about a factor 40 smaller than $M_X$.

Let us now discuss the consequences of our results for the
leptogenesis scenario \cite{lepto} (even though our findings can be
easily generalized to any out-of-equilibrium scenario for the
production the baryon asymmetry) where the lepton asymmetry ($L$) is
reprocessed into baryon number by the anomalous sphaleron transitions
\cite{kl}. Again we will assume case III for which the final
reheating temperature is fixed by the inflaton mass.

In the simplest leptogenesis scenario, the lepton asymmetry is
generated by the out-of-equilibrium decay of a massive right-handed
Majorana neutrino, whose addition to the Standard Model spectrum
breaks $B-L$.

Let us indicate by $n_{N}$ the number density per comoving volume of
the lightest right-handed neutrino $N$, the one whose final decay
(into left-handed leptons and Higgs bosons) is responsible for the
generation of the lepton asymmetry.  We can approximate the Boltzmann
equation for $N$ as
\begin{equation}
\frac{dn_N}{dt}+3H n_N=-\Gamma_N\left(n_N -
\left(n_N\right)_{eq}\right),
\end{equation} 
where $\Gamma_N$ is the decay rate of $N$ for the processes
$N\rightarrow H^\dagger \ell_L ,H{\bar \ell}_L$.  Assume again that
$M_\phi<M_N$, and that the actual density $n_N$ is much less than its
equilibrium value $(n_N)_{eq}= 2(M_N T/2\pi)^{3/2}e^{-M_N/T}$.  Since
the dominant contribution to the production of right-handed neutrinos
will come from end of reheating when the temperature is of the order of
$M_\phi$, we can estimate the ratio between the number density of
$N$-particles and the entropy density at the end of reheating to be
\begin{eqnarray}
\label{k}
\frac{n_N}{s} & \simeq & \frac{10^{-1}}{g_*^{3/2}}
	\left(\frac{\Gamma_N M_{Pl}}{M_\phi^2}\right)
	\left(\frac{M_N}{M_\phi}\right)^{3/2}e^{-M_N/M_\phi}\nonumber \\
	& \simlt & \frac{10^{-1}}{g_*} \left(\frac{M_N}{M_\phi}\right)^{3/2}
e^{-M_N/M_\phi},
\end{eqnarray}
where in the last expression we have imposed that when right-handed
neutrinos are produced, their direct decay is inefficient, {\em i.e.,}
\begin{equation}
K=\left.\frac{\Gamma_N}{H}\right|_{T\simeq M_\phi}\simeq
\frac{\Gamma_N M_{Pl}}{g_*^{1/2} M_\phi^2}
\simlt 1.
\end{equation}
The limiting case $K\sim 1$ would mean that the right-handed
neutrinos enter into chemical equilibrium as soon as they are
generated.

The ratio in Eq.\ (\ref{k}) remains constant until the right-handed
neutrinos decay generating a lepton asymmetry $L=\epsilon (n_N/s)$,
where $\epsilon$ is the small parameter containing the information
about the CP-violating phases and the loop factors.  The corresponding
baryon asymmetry is $B=(28/79)L$, assuming only Standard Model degrees
of freedom, and therefore the final baryon asymmetry is bounded to be
smaller than
\begin{equation}
\label{pp}
B=10^{-4} \epsilon \left(\frac{M_N}{M_\phi}\right)^{3/2}
e^{-M_N/M_\phi}.
\end{equation}
For a hierarchical spectrum of right-handed neutrinos, it has been
shown that that there is a model independent upper bound on the CP
asymmetry produced in the right-handed neutrino decays, $\epsilon
\simlt 3 m_{\nu_{3}} M_{N}/(8 \pi v^2)$, where $m_{\nu_{3}}$ is the
mass of the heaviest of the left-handed neutrinos and $v$ is the
Standard Model Higgs vacuum expectation value \cite{bound}. Therefore,
the maximum value of the baryon asymmetry in Eq.\ (\ref{pp}) is further
bounded from above by (taking $ m_{\nu_{3}}\sim 0.07$ eV, the
atmospheric neutrino mass scale)
\begin{equation}
B \simlt 10^{-6}\left(\frac{M_N}{10^{10}\, {\rm GeV}}\right)
\left(\frac{M_N}{M_\phi}\right)^{3/2}
e^{-M_N/M_\phi}.
\end{equation}
The requirement that $B$ is larger than $2\times 10^{-11}$ implies
that the ratio $M_N/M_\phi$ cannot be larger than about 15.

%%%%%%%%%%%%%%%%%%%%%%%%%%%%%%%%%%%%%%%%%%%%%%%%%%%%%%%%%%%%%%%%%%
%%%%%%%%%%%%%%%%%%%%%%%%%%%%%%%%%%%%%%%%%%%%%%%%%%%%%%%%%%%%%%%%%%
\section{conclusions}
%%%%%%%%%%%%%%%%%%%%%%%%%%%%%%%%%%%%%%%%%%%%%%%%%%%%%%%%%%%%%%%%%%
%%%%%%%%%%%%%%%%%%%%%%%%%%%%%%%%%%%%%%%%%%%%%%%%%%%%%%%%%%%%%%%%%%

Reheating after inflation occurs due to particle production by the
oscillating inflaton field, and its dynamics is very rich.  In this
paper we have observed that the inflaton decay products acquire plasma
masses during the reheating phase.  The plasma masses may render
inflaton decay kinematicaly forbidden, causing the temperature to
remain frozen for a period at a plateau value. This happens in any
models where the decay rate of the inflaton field $\Gamma_\phi$ is
larger than about $M_\phi^2/M_{Pl}$. This condition does not seem to
be very restrictive. If the condition is met, the final reheating
temperature is uniquely determined by the inflaton mass, and not by its
coupling. If the reheating dynamics is mainly dominated by a scalar
field $\chi$ different from the inflaton, then the final reheating
temperature may be determined in terms of the mass of the
$\chi$ field. An example is if reheating takes place along a flat
supersymmetric direction whose mass is the soft supersymmetry breaking
scale $\widetilde{m}\sim 10^2$ GeV and whose couplings to ordinary
matter is of order unity. In such a case, the effects of plasma
blocking are crucial to determine the final reheating temperature to
be $T_{RH}\sim \widetilde{m}$.

We have shown that our results are relevant for the thermal production
of dangerous relics during reheating, for extracting bounds on
particle physics models of inflation from Cosmic Microwave Background
anisotropy data, for the production of massive dark matter candidates
during reheating, and for models of baryogenesis or leptogensis where
massive particles are produced during reheating.

%%%%%%%%%%%%%%%%%%%%%%%%%%%%%%%%%%%%%%%%%%%%%%%%%%%%%%%%%%%%%%%%%%
%%%%%%%%%%%%%%%%%%%%%%%%%%%%%%%%%%%%%%%%%%%%%%%%%%%%%%%%%%%%%%%%%%
\acknowledgments{{This work was supported in part by NASA grant NAG5-10842.}
%%%%%%%%%%%%%%%%%%%%%%%%%%%%%%%%%%%%%%%%%%%%%%%%%%%%%%%%%%%%%%%%%%
%%%%%%%%%%%%%%%%%%%%%%%%%%%%%%%%%%%%%%%%%%%%%%%%%%%%%%%%%%%%%%%%%%

%%%%%%%%%%%%%%%%%%%%%%%%%%%%%%%%%%%%%%%%%%%%%%%%%%%%%%%%%%%%%%%%%%
%%%%%%%%%%%%%%%%%%%%%%%%%%%%%%%%%%%%%%%%%%%%%%%%%%%%%%%%%%%%%%%%%%

%%%%%%%%%%%%%%%%%%%%%%%%%%%%%%%%%%%%%%%%%%%%%%%%%%%%%%%%%%%%%%%%%%
%%%%%%%%%%%%%%%%%%%%%%%%%%%%%%%%%%%%%%%%%%%%%%%%%%%%%%%%%%%%%%%%%%

\begin{thebibliography}{99}
\frenchspacing
%%%%%%%%%%%%%%%%%%%%%%%%%%%%%%%%%%%%%%%%%%%%%%%%%%%%%%%%%%%%%%%%%%
%%%%%%%%%%%%%%%%%%%%%%%%%%%%%%%%%%%%%%%%%%%%%%%%%%%%%%%%%%%%%%%%%%

\bibitem{review} 
D. H. Lyth and A. Riotto,
Phys. Rept. 314, 1 (1999); 
see also A. Riotto, 
arxiv:hep-ph/0210162.

\bibitem{book} 
E. W. Kolb and M. S. Turner, 
{\it The Early Universe}, (Addison-Wesley, Menlo Park, Ca., 1990). 

\bibitem{pre} 
L. Kofman, A. D. Linde, and A. A. Starobinsky,
Phys. Rev. Lett. 73, 3195 (1994); 
L. Kofman, A. D. Linde, and A. A. Starobinsky,
Phys. Rev. D 56, 3258 (1997).

\bibitem{st} 
R. J. Scherrer and M. S. Turner, 
Phys. Rev. D 31, 681 (1985).

\bibitem{ckr} 
D. J. Chung, E. W. Kolb, and A. Riotto,
Phys. Rev. D 60, 063504 (1999). 

\bibitem{gkr} 
G. F. Giudice, E. W. Kolb, and A. Riotto, 
Phys. Rev. D 64, 023508 (2001).

\bibitem{sugra} 
H. P. Nilles,
Phys. Rept. 110, 1 (1984).

\bibitem{ellis} 
J. Ellis, J. Kim, and  D. V. Nanopoulos, 
Phys. Lett. B145, 181 (1984);
L. M. Krauss, 
Nucl. Phys. B227, 556 (1983);
M. Yu. Khlopov and A. D. Linde,
Phys. Lett. 138B, 265 (1984).	         

\bibitem{nucleo}
R. H. Cyburt, J. Ellis, B. D. Fields, and K. A. Olive,
Phys. Rev. D 67, 103521 (2003).

\bibitem{sdm}
N. Fornengo, A. Riotto, and S. Scopel,
Phys. Rev. D 67, 023514 (2003).

\bibitem{baryo}
S. Davidson, M. Losada, and A. Riotto,
Phys. Rev. Lett. 84, 4284 (2000).

\bibitem{weldon} 
H. A. Weldon,
Phys. Rev. D 26, 2789 (1982).

\bibitem{andrei} 
A. D. Linde,
Phys. Lett. B 160, 243 (1985).

\bibitem{wmap} C. L. Bennett {\it et al.},
arxiv:astro-ph/0302207; 
H. V. Peiris, {\it et. al.,}
arxiv:astro-ph/0302225.

\bibitem{barger}V. Barger, H. S. Lee, and D. Marfatia,
Phys. Lett. B 565, 33 (2003).

\bibitem{us} 
W. H. Kinney, A. Melchiorri and A. Riotto,
Phys. Rev. D 63, 023505 (2001);
W. H. Kinney, E. W. Kolb, A. Melchiorri, and A. Riotto,
arxiv:hep-ph/0305130.

\bibitem{leach} 
S. M. Leach and A. R. Liddle,
arxiv:astro-ph/0306305.

\bibitem{scottlam}
S. Dodelson and L. Hui,
arxiiv:astro-ph/0305113.

\bibitem{shdm}
D. J. Chung, E. W. Kolb, and A. Riotto,
Phys. Rev. D 59, 023501 (1999);
D. J. Chung, E. W. Kolb, and A. Riotto,
Phys. Rev. Lett. 81, 4048 (1998);
E. W. Kolb, D. J. Chung, and A. Riotto,
arxiv:hep-ph/9810361; 
D. J. Chung, P. Crotty, E. W. Kolb, and A. Riotto,
Phys. Rev. D 64, 043503 (2001).

\bibitem{lepto} M. Fukugita and T. Yanagida,
Phys. Lett. B 174, 45 (1986).

\bibitem{kr1} 
V. A. Kuzmin and V. A. Rubakov, 
Phys. Atom. Nucl. 61, 1028 (1998).

\bibitem{kr2} 
V. Berezinsky, M. Kachelriess, and A. Vilenkin, 
Phys. Rev. Lett. 79, 4302 (1997).

\bibitem{bs}
M. Birkel and S. Sarkar, Astropart. Phys. 9, 297 (1998).

\bibitem{kl} S. Yu. Khlebnikov, and M. E. Shaposhnikov, 
Nucl. Phys. B 308, 885 (1988); 
J. A. Harvey and M. S. Turner, Phys. Rev. D 42, 3344 (1990).

\bibitem{bound} 
S. Davidson and A. Ibarra,
Phys. Lett. B 535, 25 (2002).


\end{thebibliography}
\end{document}